\documentclass[aps,prl,twocolumn,superscriptaddress,10pt,floatfix,amsmath,amssymb]{revtex4-2}
\usepackage[bookmarks=true,colorlinks=true,urlcolor=blue,linkcolor=blue,citecolor=blue,breaklinks]{hyperref}

\usepackage{amsmath, amstext, amssymb, amsfonts, amsxtra}
\usepackage[pdftex]{graphicx}
\usepackage[usenames]{color}
\newcommand{\como}{Center for Nonlinear and Complex Systems, Dipartimento di Scienza e Alta Tecnologia, Universit\`a degli Studi dell'Insubria, via Valleggio 11, 22100 Como, Italy}
\newcommand{\infn}{Istituto Nazionale di Fisica Nucleare, Sezione di Milano, via Celoria 16, 20133 Milano, Italy}
\newcommand{\brazil}{International Institute of Physics, Federal University of Rio Grande do Norte, Campus Universit\'ario - Lagoa Nova, CP. 1613, Natal, Rio Grande Do Norte 59078-970, Brazil}
\newcommand{\xiamen}{Department of Physics and Fujian Provincial Key Laboratory of Low Dimensional Condensed Matter Physics, Xiamen University, Xiamen 361005, Fujian, China}
\newcommand{\lanzhou}{Lanzhou Center for Theoretical Physics, Lanzhou University, Lanzhou 730000, Gansu, China}

\begin{document}

\title{Classical-to-Quantum Transition in Blackbody Radiation Thermodynamics: \\
Planck’s Law without Statistical Assumptions}

\author{Jiao Wang}
\affiliation{\xiamen}
\affiliation{\lanzhou}
\author{Giulio Casati}
\affiliation{\como}
\affiliation{\brazil}
\author{Giuliano Benenti}
\affiliation{\como}
\affiliation{\infn}

\begin{abstract}
Inspired by Fermi’s model of radiation–matter interaction, we investigate the classical dynamics of a closed, finite-energy system, in which matter is coupled to an electromagnetic field. We find that energy is progressively transferred to increasingly high-frequency field modes, so that the effective temperature of both matter and radiation approaches zero in the long-time limit — in sharp contradiction with the finite-temperature equilibrium assumed by classical blackbody thermodynamics. Requiring consistency with the Stefan–Boltzmann and Wien laws identifies a characteristic time $t_L$, at which both laws are simultaneously satisfied. Strikingly, classical spectra obtained for different energies and cavity sizes collapse onto the universal Planck distribution at $t_L$. We conjecture that quantum dynamical localization arrests the classical energy cascade at $t_L$, and that the Planck distribution emerges as a dynamical consequence of this arrest — without invoking any a priori statistical assumption.
\end{abstract}

\maketitle

\textit{Introduction}---Why does a purely classical argument, based on thermodynamics and classical electromagnetism, correctly predict the Stefan–Boltzmann law~\cite{Boltzmann84}, even though the law contains Planck's constant and is therefore inherently quantum~\cite{Paul15}? And what is 
the long-time behavior of a finite-energy classical system in which matter interacts with the electromagnetic field? These questions, rooted in the blackbody problem that launched quantum theory, remain surprisingly open. Planck's law successfully describes thermal radiation and resolves the ultraviolet catastrophe~\cite{Feynman2011_Vol1_Ch41}, yet the dynamical connection between classical microscopic evolution, thermodynamics, and the quantum radiation law remains unclear.

For systems with a finite number of degrees of freedom, ergodicity provides the foundation for equilibrium statistical mechanics~\cite{khinchin1949,Arnold1989,Viana_Oliveira_2016}, leading to energy equipartition among quadratic degrees of freedom. 
The equipartition theorem, however, does not generalize directly to infinite-dimensional systems~\cite{Giulio83}, such as the electromagnetic field. If equipartition were realized among infinitely many field modes, conservation of energy would imply that the average energy per mode vanishes as the number of excited modes grows, corresponding to an effective temperature tending to absolute zero~\cite{Jeans16}. 
Alternatively, the system may approach a nontrivial stationary state at finite temperature, characterized by a non-equipartition distribution.
Classical blackbody thermodynamics assumes a finite-temperature equilibrium state satisfying the Stefan–Boltzmann and Wien laws. 
The key question is therefore whether the long-time dynamics of a classical system with infinitely many degrees of freedom leads to equipartition, as suggested by statistical-mechanical arguments, or to an energy distribution consistent with thermodynamic laws.
This question touches on fundamental issues in nonequilibrium statistical mechanics, ergodic theory for infinite-dimensional systems, nonlinear dynamics, and the dynamical foundations of thermalization.

Addressing the above questions requires a direct dynamical analysis of energy transfer and equilibration beyond the standard framework of finite-dimensional systems.
Here we revisit the classical blackbody problem from this perspective by generalizing a model originally introduced by Enrico Fermi~\cite{Fermi32}, in which an electromagnetic field interacts with a charged particle in a finite cavity. By numerically integrating the classical equations of motion over very long times, we find that energy is progressively transferred to higher-frequency modes. As the number of excited modes grows, the average energy per mode decreases, corresponding to an effective cooling 
of both matter and radiation toward absolute zero.
This behavior is in sharp contradiction with the finite, nonzero-temperature equilibrium state assumed in classical blackbody thermodynamics and characterized by the Stefan–Boltzmann and Wien laws.

Motivated by these findings and in an effort to provide a bridge between classical mechanics, thermodynamics, and quantum theory, we propose quantum dynamical localization~\cite{casati79, Izrailev1990, benenti2026} as a mechanism capable of suppressing the classical transfer of energy to arbitrarily high-frequency modes. Such localization would prevent the unbounded spreading of energy over field modes, thereby allowing a finite-temperature blackbody state to emerge. We argue that the Planck distribution originates from this localization mechanism and present numerical evidence supporting this scenario.

\begin{figure}
\includegraphics[width=8.0cm]{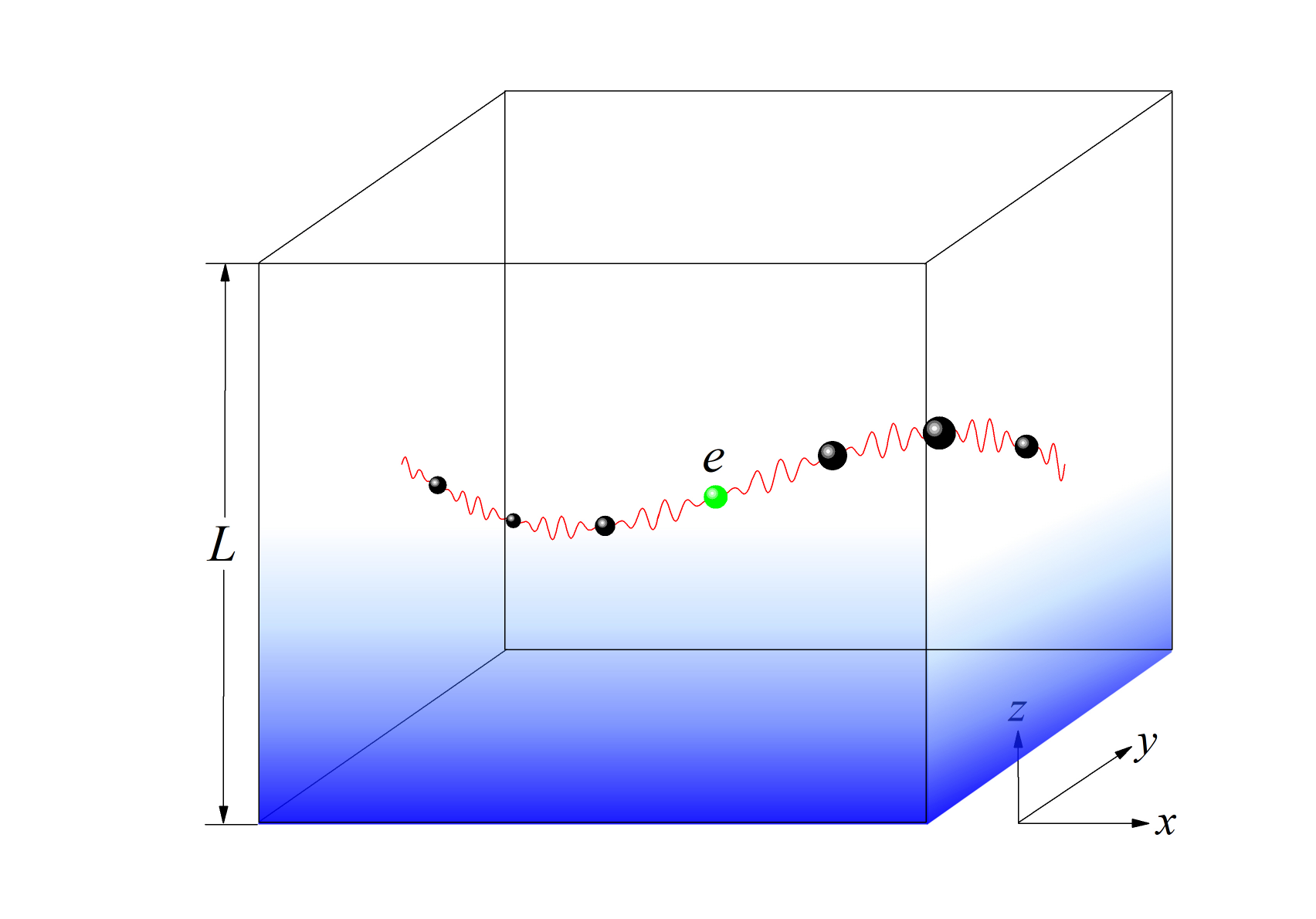}
\vskip-2mm
\caption{Schematic representation of our model. A single charged particle with charge $e$ (shown in green) mediates the exchange of energy between the electromagnetic field within a cubic cavity and a chain of $N_p-1$ neutral point particles (shown in black) interacting with their nearest neighbors. 
}
\label{fig:model}
\end{figure}

\textit{Model}---Our model, schematically illustrated in Fig. \ref{fig:model}, generalizes the one proposed by Enrico Fermi in his review on quantum electrodynamics \cite{Fermi32}. The system consists of $N_p$ interacting particles of mass $m$, only one of which carries a charge $e$. The particles are confined within a cubic cavity of side length $L$, with periodic boundary conditions
\footnote{However, if the fixed boundary conditions are adopted instead, the results are found to be the same qualitatively, suggesting that our analysis and conclusions do not depend on the particular boundary conditions, either. For the fixed boundary conditions of the cavity, it can be regarded as being enclosed by three pairs of parallel mirrors and compared with \eqref{eq:H}, the Hamiltonian of the system lacks the term $V(\mathbf{r}_{N_p}-\mathbf{r}_{1})$ and all the modes that have at least one even component $n_x$, $n_y$, and $n_z$ in their wave number vectors.}.
The motion of the charged particle excites the electromagnetic field, which in turn acts back on the particle; thus, neither the particle’s motion nor the field is prescribed {\it a priori}. The Coulomb self-interaction of the electron is neglected~\footnote{Here, we neglect the divergent electrostatic self-energy of point charges. This longstanding problem concerns the internal structure of the electron and is beyond the scope of the present work. As emphasized by Fermi~\cite{Fermi32}, radiation-theory problems not involving electron structure can be treated independently of this unresolved issue.}. The electromagnetic field is described in terms of normal mode coordinates, each mode characterized by a wave number vector
$\mathbf{n}=(n_x, n_y, n_z)$ ($n_x,n_y,n_z=0,1,2,...$), or equivalently a wave vector $\mathbf{k_n}=2\pi\mathbf{n}/L$, with frequency
$\omega_\mathbf{n}=c|\mathbf{k_n}|$, where $c$ is the speed of light.

The Hamiltonian of the system is
\begin{eqnarray}
\label{eq:H}
H=\sum_{i=2}^{N_p}[\frac{\mathbf{p}_i^2}{2m}+V(\mathbf{r}_i-\mathbf{r}_{i-1})]+V(\mathbf{r}_{N_p}-\mathbf{r}_{1})
\hskip1.5mm\nonumber\\
+\frac{1}{2m}[\mathbf{p}_1-\frac{e}{c}\sum_{\alpha,\mathbf{n},j} q_{\mathbf{n},j}^\alpha \mathbf{a}_{\mathbf{n},j}^\alpha]^2\hskip7mm\nonumber\\
+\frac{1}{2}\sum_{\alpha,\mathbf{n},j}[\frac{(p_{\mathbf{n},j}^\alpha)^2}{L^3}+L^3 \omega_\mathbf{n}^2(q_{\mathbf{n},j}^\alpha)^2].
\end{eqnarray}
Here, $\mathbf{p}_i$ and $\mathbf{r}_i$ denote, respectively, the momentum and the displacement from equilibrium of the $i$-th particle on a one-dimensional lattice. In particular, for the charged particle we set $i=1$, and write the vector potential as ${\bf A}({\bf r}_1, t)=\sum_{\alpha,\mathbf{n},j}  q_{\mathbf{n},j}^\alpha(t)  \mathbf{a}_{\mathbf{n},j}^\alpha({\bf r}_1)$.
Each mode $\mathbf{n}$ has four components, corresponding to a basis of 
even/odd functions ($\alpha=e, o$) and to polarization vectors orthogonal to $\mathbf{k_n}$ ($j=1, 2$) (see End Matter for details).
The quantities $p_{\mathbf{n},j}^\alpha$ are the momenta conjugate to the field coordinates $q_{\mathbf{n},j}^\alpha$.
Chaotic behavior is ensured by assuming the following interaction potential between two neighboring particles 
\begin{eqnarray}
\label{eq:V}
V(\mathbf{r}_i-\mathbf{r}_{i-1})=
\frac{\eta}{4}(\mathbf{r}_i-\mathbf{r}_{i-1})^4,
\end{eqnarray}
where $\mathbf{r}_i=(x_i, y_i, z_i)$, and $\eta$ sets the strength of the interaction.

\begin{figure}
\vskip-1mm
\includegraphics[width=8.8cm]{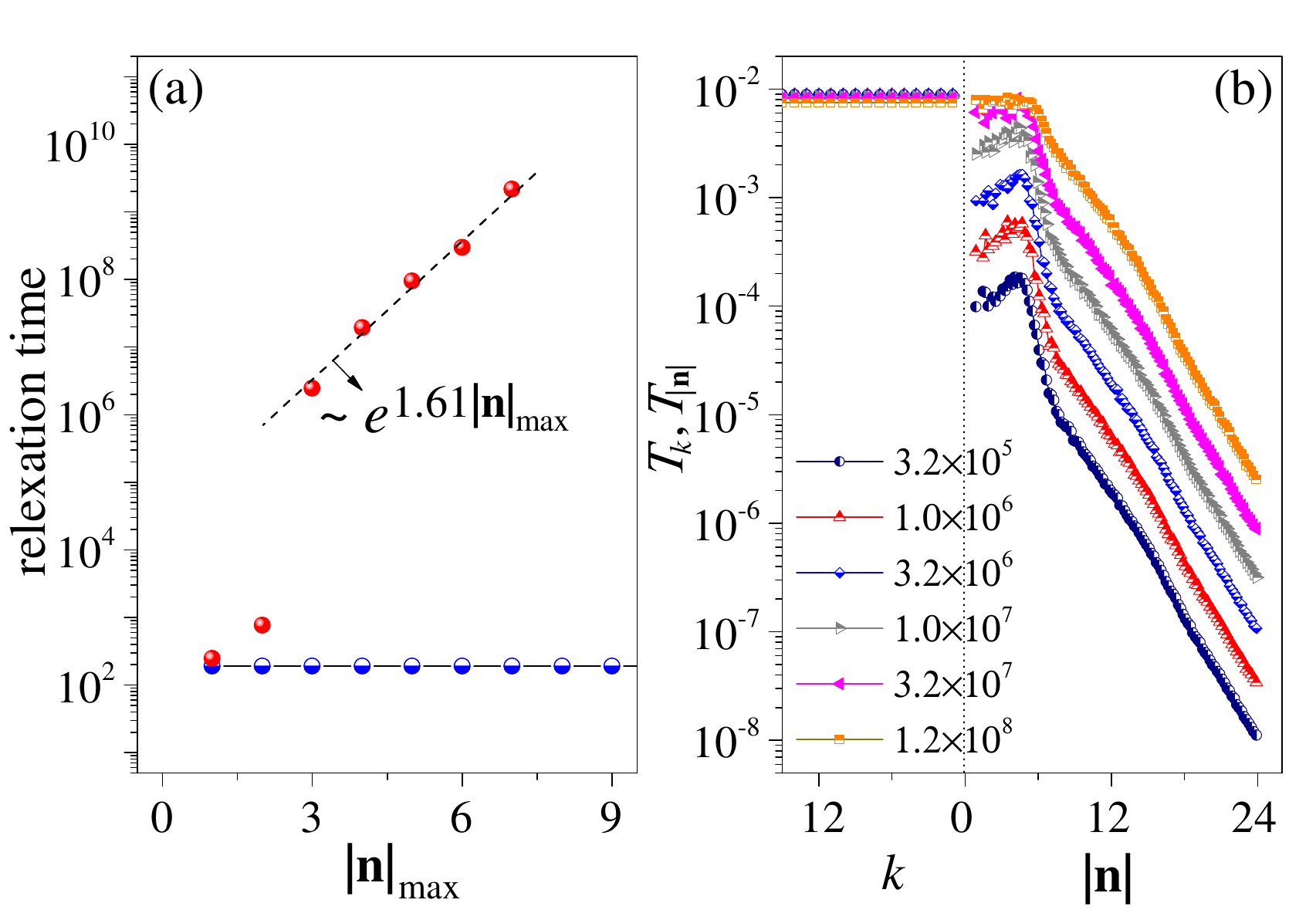}
\caption{(a) The relaxation time (defined as the time to reach equipartition within $90\%$ accuracy) for $N_p=100$, $E_{tot}=2$, and $L=20$. The blue and red data points represent the relaxation time of the matter subsystem and the whole system, respectively. The best fitting of the latter suggests that it grows exponentially as $Ce^{1.61|\mathbf{n}|_{\rm max}}$ with $C\approx e^{10}$. (b) The effective temperatures of particles (to the left of the dotted line) 
and of field modes (to the right of the dotted line) for $N_p=4000$, $E_{tot}=80$, and $L=40$, at various times from $t=3.2\times 10^5$ to $1.2\times 10^8$. 
}
\label{fig:relax-TkTn}
\end{figure}

The second line of the r.h.s. of Eq. \eqref{eq:H} describes the coupling between the charged particle and the electromagnetic field. When the charge vanishes ($e=0$), field and matter are decoupled, and each field mode reduces to an independent harmonic oscillator. The motion of particles is chaotic for $N_p\ge 4$, regardless of how small their energy $E_p$ is. Indeed, owing to the scaling property of the particle chain
\footnote{The dynamics of the matter lattice model at different values of $E_p$, $m$, and $\eta$ are equivalent, as can be verified straightforwardly
by rescaling 
$\tilde{\mathbf{r}}_i=\mathbf{r}_i(\eta/ E_p)^{1/4}$, $\tilde{\mathbf{p}}_i=\mathbf{p}_i (m E_p)^{-1/2}$,
and $\tilde{t}= t (\eta E_p)^{1/4} m^{-1/2}$.},  
it can be shown that all its $6N_p$ Lyapunov exponents scale as $\lambda_l\sim (\eta E_p)^{1/4}m^{-1/2}$, $l=1, \cdots, 6N_p$. The power spectrum exhibits the same scaling behavior. 

The equations of motion derived from the Hamiltonian \eqref{eq:H} are numerically integrated using a leapfrog-like algorithm \cite{Galgani89}, which is more accurate than the standard Runge-Kutta method.
Initially, all field modes are unexcited and all particles are placed at their equilibrium positions. The charged particle is at rest, while the neutral particles are assigned random velocities such that their total momentum vanishes and their total kinetic energy equals the total energy of the entire system, denoted by $E_{tot}$ \footnote{To reduce fluctuations in the computed quantities, we perform an ensemble average over eight realizations with different random initial conditions together with a partial time average over the time interval $(0.5t, 1.5t)$ for quantities at time $t$.}.

In our simulations, we use dimensionless units with
$c=1$, $k_{\rm B}=1$, and $4\pi\epsilon_0=1$, where $\epsilon_0$ is the vacuum permittivity. For the charged particle we take one electron for which we set $m=1$ and $e=0.01$. All computations reported in this work are classical and therefore do not explicitly involve Planck's constant $\hbar$. However, once the dimensionless charge $e$ is fixed, the value of  $\hbar$ relevant for comparison with quantum scales is determined by the fine-structure constant $\alpha=\frac{e^2}{\hbar c}\simeq\frac{1}{137}$  with $4\pi\epsilon_0=1$. Thus for  
$e=0.01$ and $c=1$, we obtain $\hbar\approx0.0137$. 
The choice of simulation parameters is constrained  by both physical and computational considerations. 
First, since $c=1$, the nonrelativistic treatment of the matter degrees of freedom requires the particles' characteristic velocity $v$ to satisfy $v\lesssim 0.1$, a conservative threshold at which relativistic corrections are of order $(v/c)^2\lesssim10^{-2}$; using
$m=k_{\rm B}=1$, so that
$T\sim v^2$, gives the approximate upper bound $T \lesssim 10^{-2}$.
Second, in numerical simulations the cavity size $L$ cannot be arbitrarily large. Since the number of field modes below a given frequency $\omega$ grows as $\omega^3$, a larger $L$ requires tracking more modes to resolve the same frequency range; at the same time, the matter--field coupling decreases with cavity volume as $L^{-3}$, so longer integration times are needed. Together, these effects make $L>40$ prohibitively demanding for our numerical simulations.
A lower bound on the temperature follows from requiring the peak of the blackbody spectrum to lie above the lowest cavity frequency $\omega_{\rm min}=2\pi c/L$. 
According to Wien's law, the peak of the Planck distribution is located at
$\omega_{\rm W} \simeq 2.82 k_{\rm B} T/\hbar$. The condition $\omega_{\rm W}>\omega_{\rm min}$ therefore yields $T \gtrsim \pi c \hbar/(1.41 k_{\rm B} L)$, which 
for $L=20$ gives $T \gtrsim 0.0015$. 
The physically relevant and numerically accessible parameter range is thus approximately $10^{-3}\lesssim T \lesssim 10^{-2}$,
$20\lesssim L \lesssim 40$. The remaining parameter $\eta$ has no significant effect on the dynamics considered here and we therefore set $\eta=1$ throughout.

\textit{Thermalization times}---A crucial aspect of the numerical analysis is the choice of the number of modes retained in computations, determined by $|\mathbf{n}|_{\rm max}$, the maximum value of $|\mathbf{n}|$. 
Since the system is ergodic and mixing, and the total number of particles in our simulations is necessarily finite, for any finite-mode truncation equipartition is eventually reached. The characteristic behaviour of this thermalization process is shown in Fig.~\ref{fig:relax-TkTn}(a). Notice, however, that while the particles rapidly thermalize among themselves, the time required to reach global equipartition, including the field modes, increases exponentially with $|\mathbf{n}|_{\max}$. The thermalization of the field is examined in more detail in Fig. \ref{fig:relax-TkTn}(b), which shows 
particles and field-mode temperatures,
$T_k$ and $T_{|\mathbf{n}|}$ (see End Matter). 
The low-frequency modes thermalize first, forming an approximately flat plateau in the mode-energy distribution that progressively extends toward higher frequencies, followed by an exponentially decaying tail. The plateau therefore provides an operational definition of an effective field temperature. The temperature associated with the plateau gradually approaches the particle temperature, while the individual mode temperatures 
$T_{|\mathbf{n}|}$ exhibit an exponentially decaying tail at high frequencies.

\textit{Effective cooling}---The main obstacle in developing a statistical 
theory of field-matter interaction lies in the fact that 
the number of field modes is infinite and the two limits $|\mathbf{n}|_{\rm max}\to \infty$ and the time $t\to \infty$ do not commute. 
The physically relevant limit requires taking
$|\mathbf{n}|_{\rm max}\to \infty$ \textit{before} $t\to \infty$. 
Reversing the order yields equipartition by ergodicity, but this is an artefact of the truncation, not a property of the infinite-dimensional system. 
In practice,  $|\mathbf{n}|_{\rm max}$ must be chosen sufficiently
large that the neglected higher-frequency modes do not affect the dynamics over the integration times considered.
Fig. \ref{fig:relax-TkTn}(b) demonstrates that a cutoff $|\mathbf{n}|_{\rm max}=24$  is sufficient for the parameters considered here
\footnote{In our computations with $|\mathbf{n}|_{\rm max}=24$ we cannot exceed the time $t=1.2 \times 10^8$. Indeed, a single run up to this time with $|\mathbf{n}|_{\rm max}=24$ requires approximately 3 months on a mainstream workstation computer with a base frequency of 2.6 GHz.}.

\begin{figure}[t!]
\includegraphics[width=8.8cm]{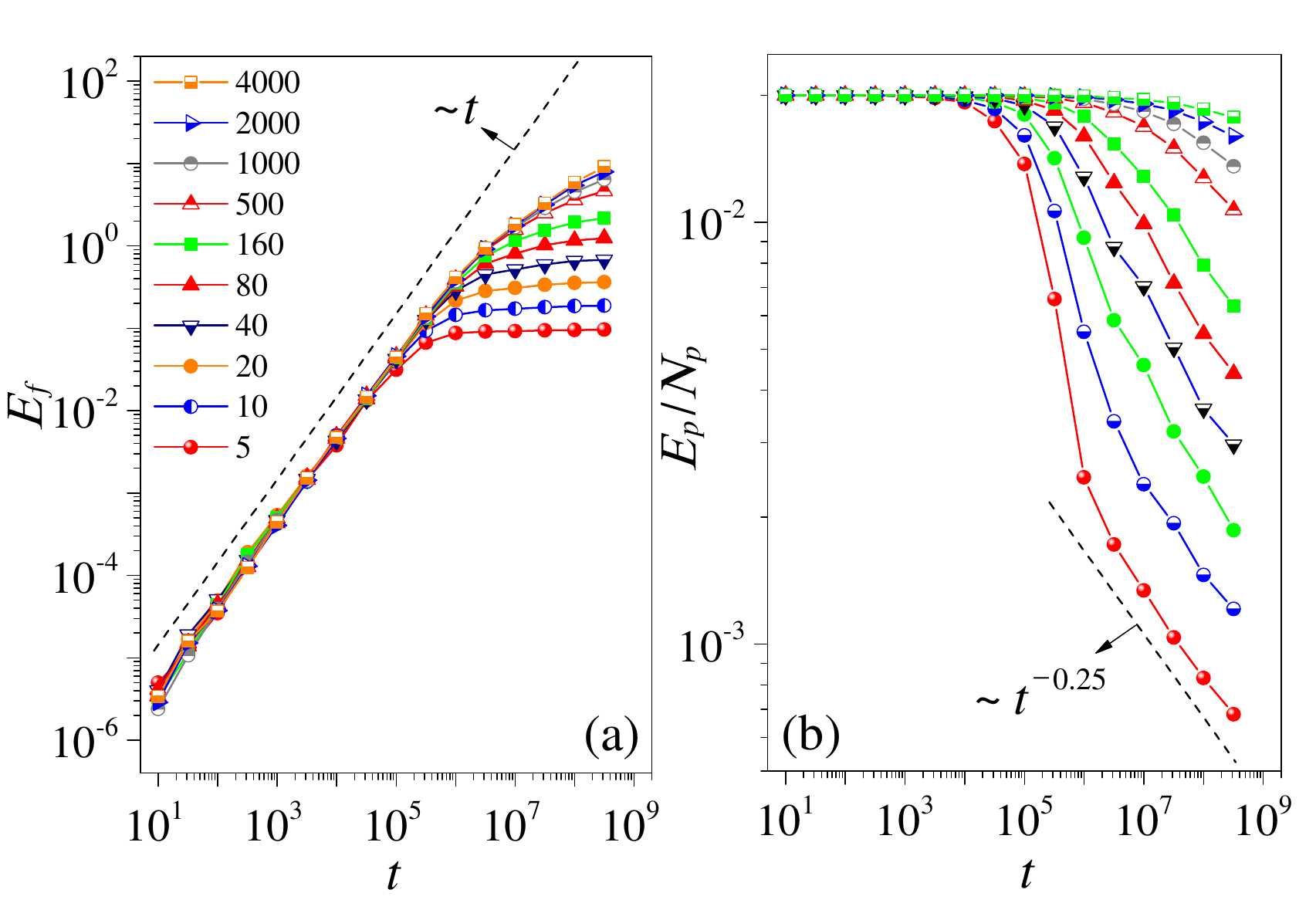}
\caption{Total field energy (a) and the average energy per particle (b) as a function of time.  Curves correspond to $N_p=5, 10, \cdots, 4000$. Here  $L=20$ and the conserved total energy is  $E_{tot}= 0.02 N_p$. Legend for (a) also applies to (b).}
\label{fig:EfEp}
\end{figure}

Figure \ref{fig:EfEp}(a) shows the total field energy $E_f$, corresponding to the third line on the right-hand side of Eq. \eqref{eq:H} as a function of time. 
Initially $E_f$ grows linearly; at later times, the growth rate decreases and $E_f$ gradually approaches the conserved total energy $E_{tot}$. This behavior is robust across the range of parameters explored and is independent of the  number of particles.
Although the relaxation time becomes progressively longer as the number of particles increases, the field energy ultimately approaches $E_{tot}$, 
independently of the amount of energy initially stored in the matter.
Correspondingly, the average energy per particle (temperature) goes to zero, as shown in Fig.~\ref{fig:EfEp}(b).
This provides a clear answer to one of the central questions addressed in this work: given sufficient time, the energy initially stored in the matter degrees of freedom is transferred to an ever-increasing number of electromagnetic field modes. Remarkably, this result provides a different perspective on the ultraviolet catastrophe. In a closed system with finite total energy, the progressive excitation of higher-frequency modes does not lead to an ultraviolet divergence; instead, the finite available energy is distributed over an increasing number of degrees of freedom, causing the effective temperature of both matter and radiation to approach absolute zero.

\begin{figure}[t!]
\vskip0.35mm
\includegraphics[width=8.2cm]{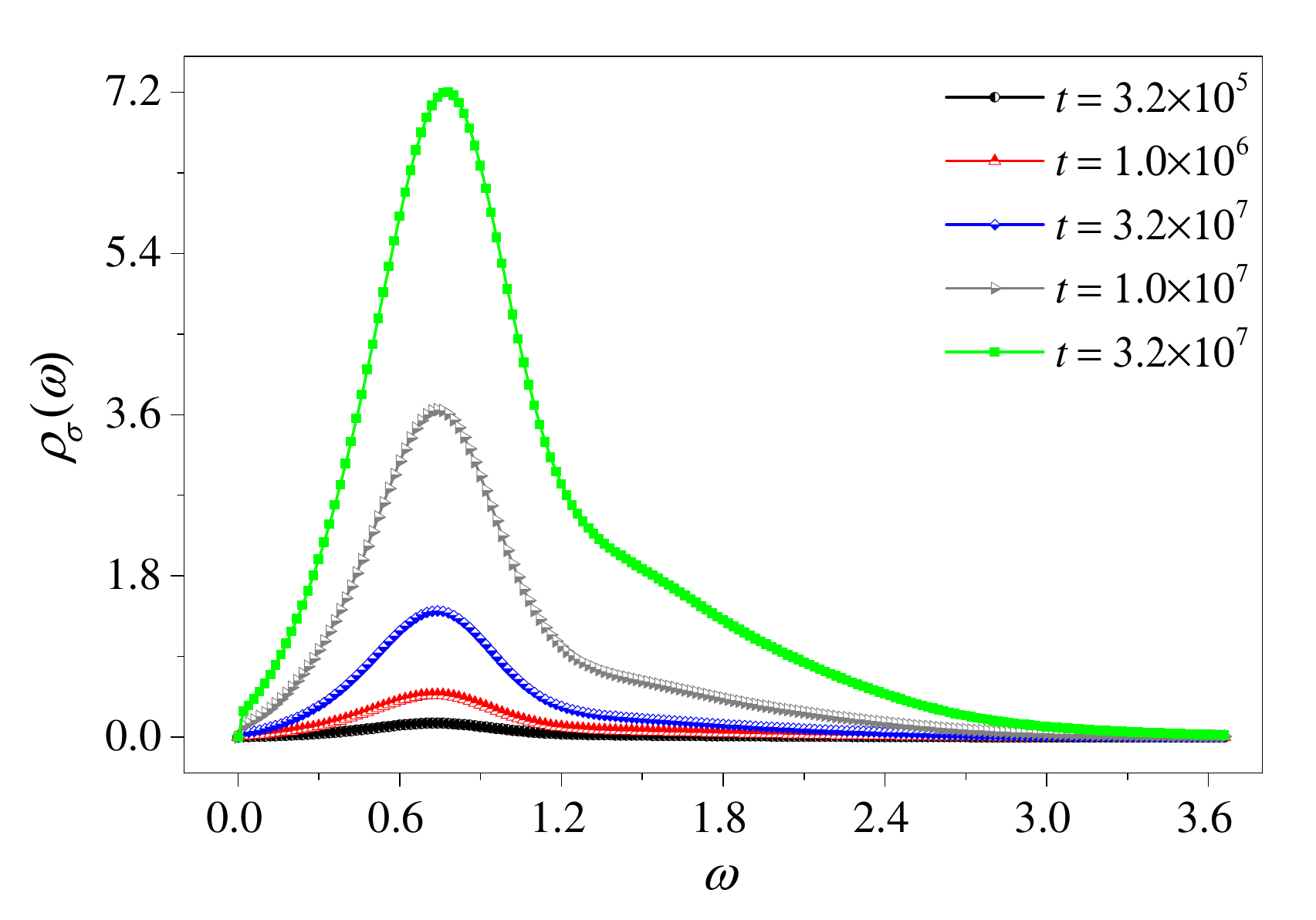}
\caption{Energy density distribution of the field modes for $E_{tot}=80$ and $L=40$, with Gaussian coarse‑grain factor $\sigma=0.02$. Here and in all subsequent figures, $N_p=4000$.}
\label{fig:rho-sigma}
\end{figure}

\textit{Classical blackbody thermodynamics}---We now compare
classical dynamics with the predictions of classical thermodynamics.
To this end we need to determine the frequency at which the energy-density spectrum reaches its maximum, in addition to the effective field temperature already introduced.
Fig. \ref{fig:rho-sigma} shows the spectral energy-density $\rho_\sigma (\omega, t)$ 
\footnote{Since the mode frequencies are discrete, $\rho(\omega, t)$ is given by a series of $\delta$ functions,
\begin{eqnarray}
\label{eq:rho}
\rho (\omega, t)= \frac{1}{2}\sum_{\mathbf{n}}\delta (\omega-\omega_{\mathbf{n}})\sum_{\alpha,j}[\frac{(p_{\mathbf{n},j}^\alpha)^2}{L^3}+L^3 \omega_\mathbf{n}^2(q_{\mathbf{n},j}^\alpha)^2].
\end{eqnarray}
For visualization and analysis purposes, it is convenient to introduce a coarse-grained spectral density. We define
\begin{eqnarray}
\label{eq:rho-sigma}
\rho_\sigma (\omega, t)=\frac{1}{\sigma \sqrt{\pi}}\int \rho(\omega',t)e^{-\frac{(\omega-\omega')^2}{\sigma^2}}d\omega',
\end{eqnarray}
which corresponds to a Gaussian coarse graining of width $\sigma$.} at different times.
Remarkably $\rho_\sigma (\omega, t)$ retains a well defined bell-shaped profile throughout the evolution. This allows us to unambiguously determine the peak frequency $\omega_{\rm peak}$ of the  classical spectrum.

\begin{figure}[t!]
\includegraphics[width=8.8cm]{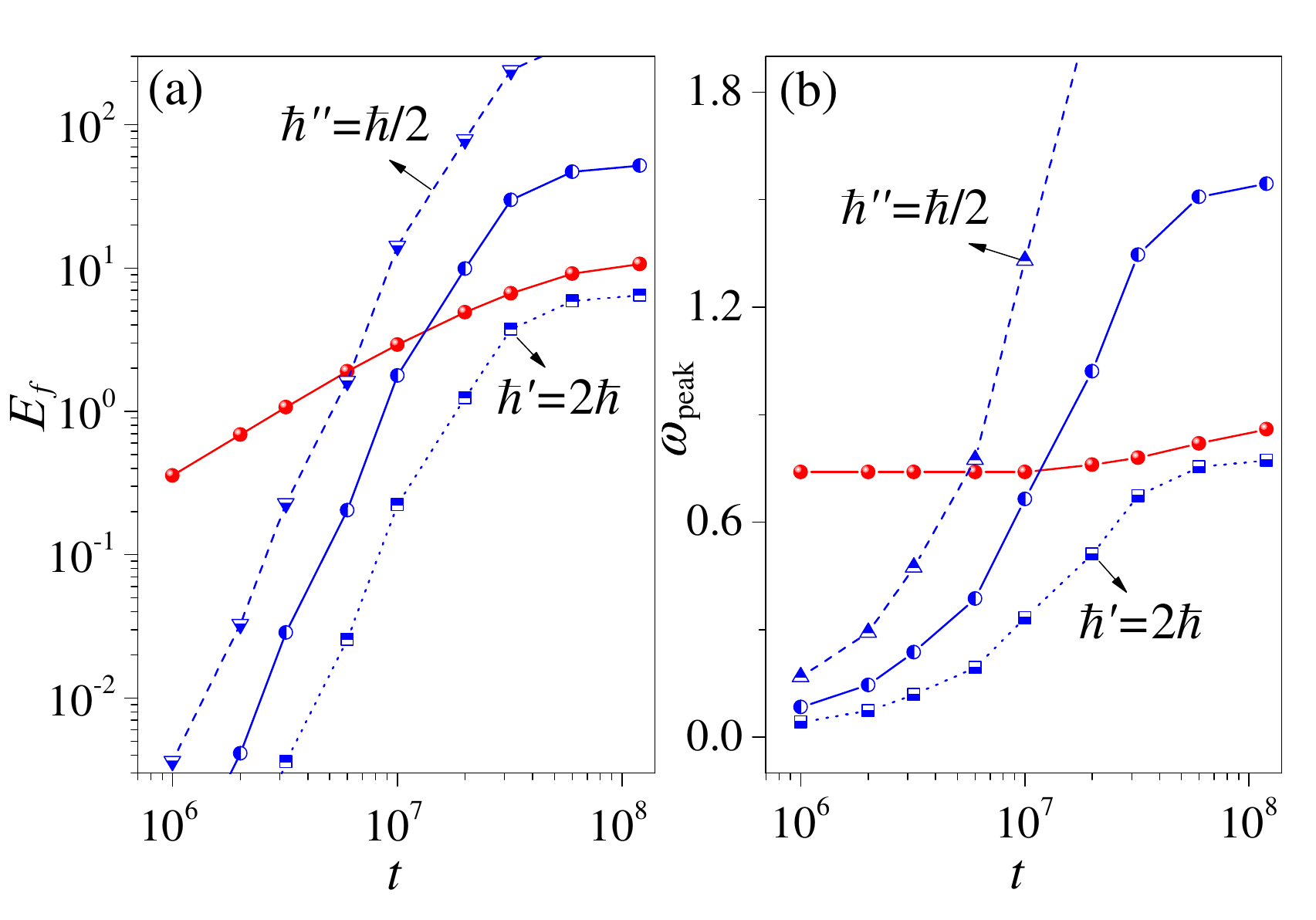}
\caption{(a) Comparison between the numerically computed field energy (red dots) and the prediction of the Stefan–Boltzmann law (blue dots) based on the field temperature at a given time. 
Blue triangles and squares show the Stefan–Boltzmann predictions obtained using, respectively, half and twice the value of Planck's constant, highlighting the sensitivity of our results to its value. 
(b) The same as (a) but for the frequency at which the energy distribution peaks. The blue curves represent the predictions by Wien's displacement law.}
\label{fig:S-B-Wien}
\end{figure}

Having determined both the effective field temperature $T$
of the low-frequency plateau and the spectral peak frequency as functions of time, we can now compare the classical dynamics with the Stefan–Boltzmann and Wien laws.
Fig.~\ref{fig:S-B-Wien}(a) compares the numerically computed field energy $E_{f}$ with the Stefan–Boltzmann prediction,
$E_{\rm {SB}}=a L^3T^4$, where $a=\frac{\pi^{2}k_{\rm B}^{4}}{15\, c^{3}\hbar^{3}}$.
The two quantities become equal at a time slightly larger than $t\sim 10^7$, indicating that the classical field distribution satisfies the Stefan–Boltzmann law at this moment of time. 
Independently, Fig.~\ref{fig:S-B-Wien}(b) compares the numerically determined frequency $\omega_{\rm peak}$ with Wien's prediction 
$\omega_{\rm W}\simeq {2.82 k_{\rm B} T}/{\hbar}$.
Remarkably, these two quantities also become equal at essentially the same time.
Thus, the Stefan--Boltzmann and Wien laws---which probe 
different properties of the radiation field---independently identify the
same characteristic time, which we denote $t_{L}$.

\begin{figure}[t!]
\vskip0.35mm
\includegraphics[width=8.2cm]{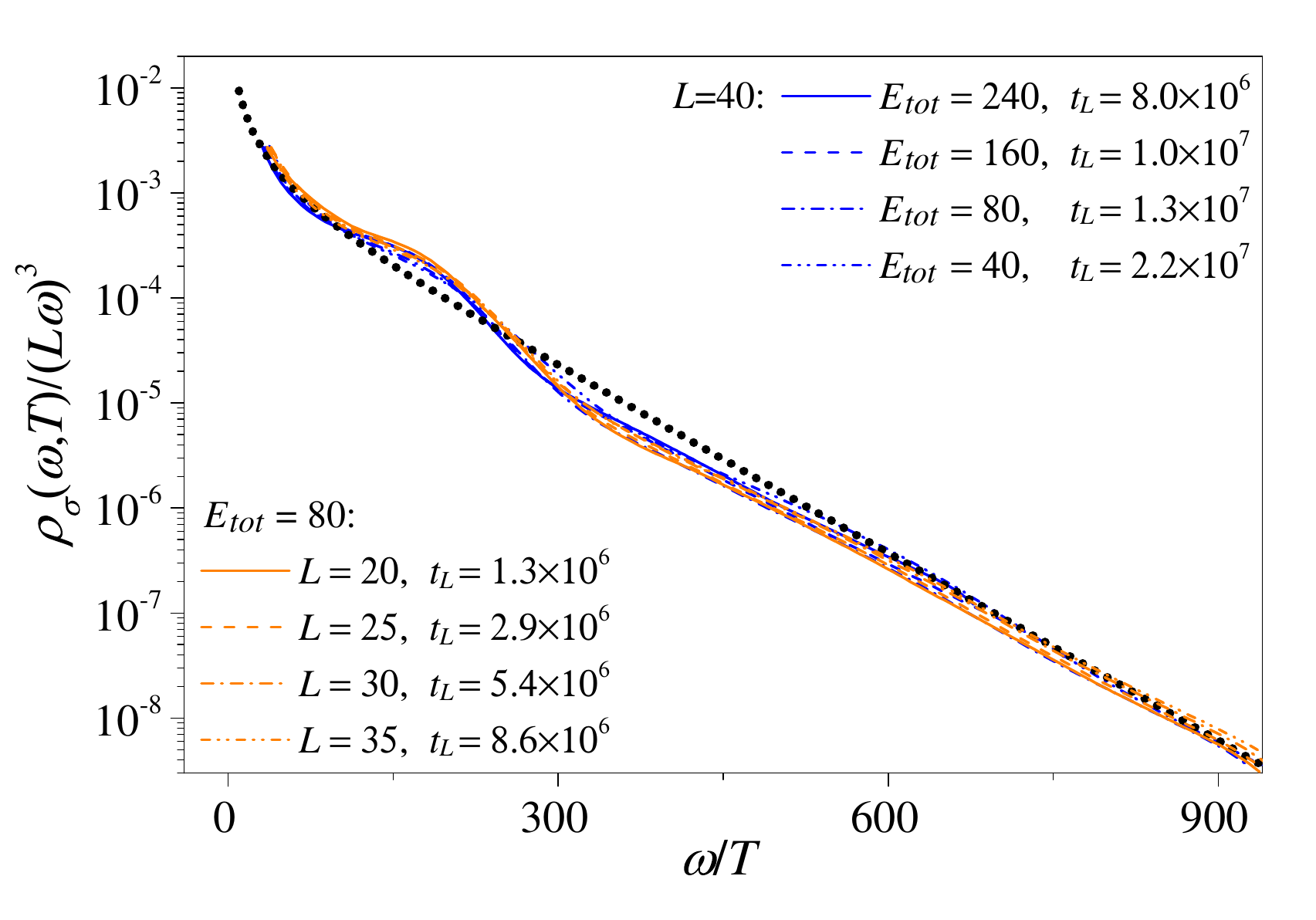}
\caption{Energy distribution of the radiation field at the localization time, $t_L$, estimated when the classical system exhibits the Wien and Stefan–Boltzmann behaviors, for various values of $E_{tot}$ at $L=40$ and various values of $L$ with $E_{tot}=80$. Here $T$ denotes the plateau temperature at time $t_L$ and the black dotted curve is for the universal function \eqref{eq:f} given by the Planck distribution.}
\label{fig:Planck}
\end{figure}

In Fig.~\ref{fig:Planck} we plot the numerically computed $\rho_\sigma(\omega,T)/(L\omega)^{3}$
as a function of the scaled frequency $x=\omega/T$, evaluated at
$t=t_{L}$ for different values of the total energy $E_{tot}$ and of the
cavity size $L$. The thick black dotted curve represents the universal
function
\begin{eqnarray}
\label{eq:f}
f(x)=\frac{\hbar}{\pi^{2}c^{3}}\frac{1}{e^{\hbar x/k_{\rm B}}-1},
\end{eqnarray}
given by the Planck distribution.
Despite the different values of $E_{tot}$, $L$, $T$, and $t_{L}$, all
numerical data approximately collapse onto the same universal curve.
Thus, at the common time selected independently by the Stefan-Boltzmann and Wien laws, the classical radiation spectrum agrees with the Planck distribution. This collapse is the central numerical result of this work.

\textit{Quantum-to-classical correspondence and localization
conjecture}---The data collapse described above calls for an explanation.
Why should the classical dynamics, frozen at a time fixed by thermodynamic consistency, reproduce the full quantum spectrum?
We propose that the answer lies in \emph{quantum dynamical
localization}~\cite{casati79}: the suppression, by quantum interference, of the 
classical diffusive energy spreading, which naturally identifies $t_{L}$ as the localization time---the timescale beyond which quantum coherence arrests the classical energy cascade.
This localization time scales as $t_L \propto E_{\rm tot}^{-1/2} L^3$, 
consistent with the strength of the matter--field coupling (see End Matter).
First identified in the paradigmatic kicked rotator, dynamical localization has subsequently been observed in a variety of driven systems ~\cite{Bayfield1989, Graham1992, Moore1995, Casati1989, Chabe2008, Benenti2003, Henry2006, Pizzamiglio2021}. It also occurs in the kicked harmonic oscillator, although no general quantitative theory of the localization time is currently available in that case~\cite{Shepelyansky1992, Jiang1993, Borgonovi1995}.

The blackbody model considered here is  
more intricate, since each field mode interacts with the charged particle, which is itself part of a deterministic chaotic system. Nevertheless, the progressive 
spreading of energy toward higher-frequency modes observed in the classical dynamics suggests that
the system may be susceptible to quantum dynamical localization. We conjecture that, in accordance with the correspondence principle, the quantum evolution initially follows the classical dynamics up to a characteristic  
localization time, beyond which quantum interference suppresses further diffusion toward higher-frequency modes. The classical energy cascade is thereby arrested, preventing the complete depletion of the matter energy and allowing a stationary state at finite temperature to emerge. 
The Planck distribution then emerges not from an
\emph{a priori} statistical assumption, but as a dynamical consequence of this localization. 
A microscopic theory of   
the localization time for the present system is beyond current reach.
However, the data collapse of Fig.~\ref{fig:Planck}, obtained
with $t_{L}$ fixed independently by the Stefan-Boltzmann and Wien laws, constitutes the numerical evidence for this conjecture.

\textit{Conclusions}---We have investigated the nonequilibrium relaxation dynamics of a three-dimensional classical blackbody radiation model. Energy is progressively transferred to increasingly high-frequency field modes, cooling both matter and radiation toward zero temperature. Thus, in the long-time limit, the classical dynamics fails to approach the finite-temperature equilibrium state described by the Stefan–Boltzmann and Wien laws.
We propose quantum dynamical localization as the mechanism that arrests the classical diffusion in mode space after a characteristic localization time $t_L$, 
independently fixed by consistency with these two laws.  
At this time, the classical spectra obtained for different total energies and cavity sizes approximately collapse onto the universal Planck distribution. 
We stress that the Planck distribution is not imposed in determining $t_L$, but emerges as an independent consequence of arresting the classical evolution at this characteristic time. Our numerical results therefore support 
the proposed localization scenario and suggest a dynamical mechanism by which quantum mechanics reconciles the underlying dynamics with finite-temperature thermodynamics.
Within this picture, the emergence of the Planck distribution appears as a consequence of dynamical localization, without invoking \emph{a priori} statistical assumptions. 
This dynamical picture may also shed light on the longstanding puzzle of how classical thermodynamics and classical electromagnetism can lead to a radiation law that is inherently quantum.
Whether the localization time $t_L$ identified here is related to known many-body localization phenomena~\cite{Huse2015,Abanin2019} remains an open question.
More broadly, our results suggest a possible connection between quantum dynamical localization, ergodicity breaking~\cite{Chushun}, and the classical-to-quantum transition in infinite-dimensional systems.

\textit{Acknowledgments.}---We acknowledge support by the National Key Research and Development Program of China (Grant No. 2023YEA1407100), the National Natural Science Foundation of China (Grants No. 12475038, No. 12247106, and No. 12247101), and the INFN through the project QUANTUM. 
We also gratefully acknowledge useful discussions with Italo Guarneri.
\bibliography{references}

\clearpage
\onecolumngrid
\appendix

\section*{End Matter}    
\twocolumngrid

\setcounter{equation}{0}
\renewcommand{\thesection}{A}
\textit{Appendix A: Details on the model}---The vector potential ${\bf A}({\bf r}_1, t)$
is expanded in the vector basis $\mathbf{a}_{\mathbf{n},j}^\alpha({\bf r}_1)$,
whose elements are eigenfunctions of the Laplacian:
\begin{equation}
\nabla^2 \mathbf{a}_{\mathbf{n},j}^\alpha({\bf r}_1)= -|\mathbf{k_n}|^2  \mathbf{a}_{\mathbf{n},j}^\alpha({\bf r}_1).
\end{equation}
Imposing the transversality condition 
\begin{equation}
\nabla\cdot \mathbf{a}_{\mathbf{n},j}^\alpha({\bf r}_1)=0,
\end{equation}
together with periodic boundary conditions in a cubic cavity of side length $L$, yields the wave vectors $\mathbf{k_n}=(2\pi/L)\mathbf{n}$, where
$\mathbf{n}=(n_x, n_y, n_z)$, with $n_x,n_y,n_z$ nonnegative integers, and a basis of even and odd eigenfunctions:
\begin{eqnarray}
\label{eq:a}
\mathbf{a}_{\mathbf{n},j}^e(\mathbf{r}_1)=c\sqrt{8\pi}\cos({\mathbf{k}_\mathbf{n}}\cdot {\mathbf{r}_1})\mathbf{u}_{\mathbf{n},j},\nonumber\\
\mathbf{a}_{\mathbf{n},j}^o(\mathbf{r}_1)=c\sqrt{8\pi}\sin({\mathbf{k}_\mathbf{n}}\cdot {\mathbf{r}_1})\mathbf{u}_{\mathbf{n},j},\hskip1mm
\end{eqnarray}
with $\mathbf{u}_{\mathbf{n},j}$ ($j=1, 2$) being two unit polarization vectors orthogonal to $\mathbf{k}_{\mathbf{n}}$ \cite{Galgani89}. 

From Hamiltonian (\ref{eq:H}) we obtain the equations of motion:
\begin{equation}
\ddot{q}_{\mathbf{n},j}^{e}+\omega_{\mathbf{n}}^{2} q_{\mathbf{n},j}^{e}
=
\frac{e \sqrt{8\pi}}{L^{3}} \dot{\mathbf{r}}_1 \cdot \mathbf{u}_{\mathbf{n},j} \cos\left(\mathbf{k}_\mathbf{n} \cdot \mathbf{r}_1\right),
\label{eq:couplinge}
\end{equation}
\begin{equation}
\ddot{q}_{\mathbf{n},j}^{o}+\omega_{\mathbf{n}}^{2} q_{\mathbf{n},j}^{o}
=
\frac{e \sqrt{8\pi}}{L^{3}} \dot{\mathbf{r}}_1 \cdot \mathbf{u}_{\mathbf{n},j} \sin\left(\mathbf{k}_\mathbf{n} \cdot \mathbf{r}_1\right),
\label{eq:couplingo}
\end{equation}

\begin{equation}
\begin{aligned}
m\ddot{\mathbf{r}}_1
&=
\eta[\left(\mathbf{r}_2-\mathbf{r}_1\right)^3-\left(\mathbf{r}_{1}-\mathbf{r}_{N_p}\right)^3]\\
&\quad
-e\sqrt{8\pi} \sum_{\mathbf{n},j} \mathbf{u}_{\mathbf{n},j}
\left[
\dot{q}_{\mathbf{n},j}^{e} \cos\left(\mathbf{k}_\mathbf{n} \cdot \mathbf{r}_1\right) 
+\dot{q}_{\mathbf{n},j}^{o} \sin\left(\mathbf{k}_\mathbf{n} \cdot \mathbf{r}_1\right)
\right]
\\
&\quad
+e\sqrt{8\pi}\,\dot{\mathbf{r}}_1 \times
\sum_{\mathbf{n},j}\left(\mathbf{k}_\mathbf{n} \times \mathbf{u}_{\mathbf{n},j}\right)
\left[
q_{\mathbf{n},j}^{o}\cos\left(\mathbf{k}_\mathbf{n}\cdot \mathbf{r}_1\right)\right.\\
&\quad
-q_{\mathbf{n},j}^{e} \sin\left(\mathbf{k}_\mathbf{n} \cdot \mathbf{r}_1\right)],
\end{aligned}
\label{eq:mr1}
\end{equation}
\begin{equation}
m\ddot{\mathbf{r}}_{N_p}
=
\eta[\left(\mathbf{r}_1-\mathbf{r}_{N_p}\right)^3-\left(\mathbf{r}_{N_p}-\mathbf{r}_{N_{p-1}}\right)^3],
\label{eq:mrNp}
\end{equation}
and
\begin{equation}
m\ddot{\mathbf{r}}_i
=
\eta [ \left(\mathbf{r}_{i+1}-\mathbf{r}_i\right)^3
-\left(\mathbf{r}_i-\mathbf{r}_{i-1}\right)^3]
\label{eq:mri}
\end{equation}
for $i=2,\dots,N_{p-1}$, with $\left(\mathbf{r}_i-\mathbf{r}_j\right)^3\equiv \left(\mathbf{r}_i-\mathbf{r}_j\right)^2\left(\mathbf{r}_i-\mathbf{r}_j\right)$ in Eqs. (\ref{eq:mr1}-\ref{eq:mri}).

\renewcommand{\thesection}{B}
\textit{Appendix B: Operational definition of temperature}---Our numerical simulations show that, after a relatively short relaxation time, the kinetic energy becomes isotropically distributed among the three spatial directions and all particles acquire approximately the same average kinetic energy. We can therefore associate a temperature $T_k$ with each particle through the equipartition relation $\langle |\mathbf{p}_k|^2/m\rangle =3k_B T_k$, where $\langle \cdot \rangle$ denotes the time average introduced in the main text. 

Similarly, for a mode $\mathbf{n}$, its four quadratic terms in the Hamiltonian, i.e., $(p_{\mathbf{n},j}^\alpha)^2/{L^3}$ with $\alpha=o, e$ and $j=1, 2$, converge on average to a value which is shared by all the modes with the same frequency. Therefore, we can define
\begin{eqnarray}
\label{eq:Tn}
T_{|\mathbf{n}|}(t)=\frac{1}{4k_B L^3 N_{|\mathbf{n}|}}\sum_{\alpha, \mathbf{n},j}\langle (p_{\mathbf{n},j}^\alpha)^2 \rangle
\end{eqnarray}
as a virtual temperature of the modes with the same frequency or the same value of $|\mathbf{n}|$ (note that $\omega_{\mathbf{n}}=2\pi c |\mathbf{n}|/L$), where $N_{|\mathbf{n}|}$ is the total number of the modes with the same $|\mathbf{n}|$.

We have also verified that replacing the momentum variables $p_{\mathbf{n},j}^\alpha$ in the right-hand side of Eq.~\eqref{eq:Tn} with the corresponding coordinate variables $q_{\mathbf{n},j}^\alpha$ yields virtually identical results. The corresponding temperatures, denoted by $\tilde T_{|\mathbf{n}|}$, are indistinguishable from $T_{|\mathbf{n}|}$ within numerical uncertainties. Therefore, the same conclusions hold when the total energy of a mode is considered, confirming the robustness of the properties exhibited by $T_{|\mathbf{n}|}$.

\begin{figure}[h]
\includegraphics[width=8.9cm]{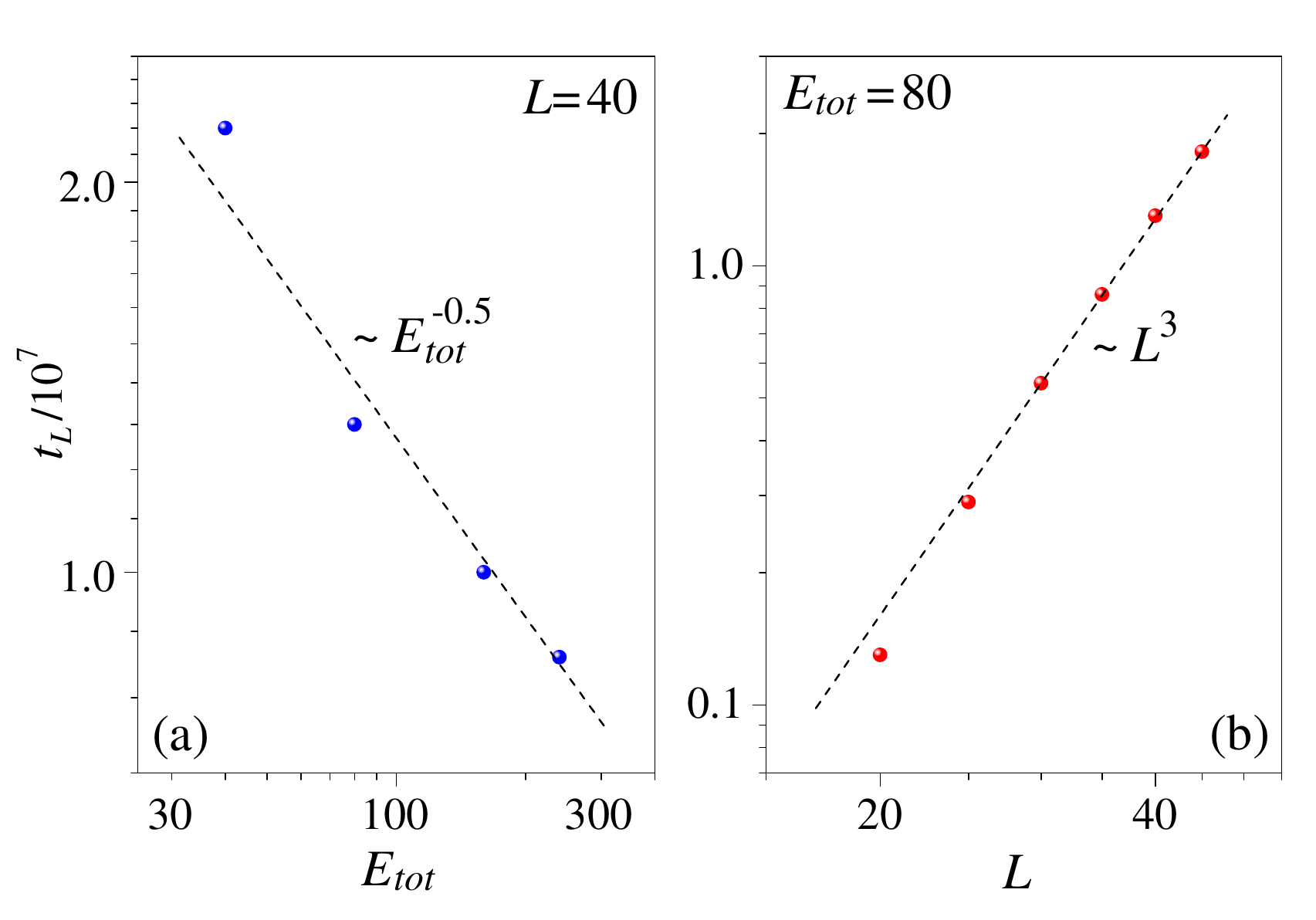}
\caption{The localization time versus the total energy $E_{tot}$ for $L=40$ (a) 
and versus $L$ for $E_{tot}=80$ (b), respectively. The dashed lines represent the scalings conjectured based on the matter-field interaction.}
\label{fig:tLvE}
\end{figure}

\renewcommand{\thesection}{B}
\textit{Appendix C: Scaling of the localization time}---In Fig.~\ref{fig:tLvE} we show the numerically computed localization time as a function of the total 
energy and of the system size. 
The results are in good agreement with the scalings 
$t_L\propto E_{tot}^{-1/2}$ and $t_L\propto L^3$, suggested
by the form of the matter-field interaction; see 
Eqs. (\ref{eq:couplinge}) and (\ref{eq:couplingo}). 

\end{document}